\documentclass{iopart}
\usepackage{graphicx}
\usepackage{latexsym}
\usepackage{amssymb}
\usepackage{verbatim}
\usepackage{iopams}

\newcommand{\chem}[1]{\ensuremath{\mathrm{#1}}}

\begin{document}

\title[Heat capacity measurements on FeAs-based compounds]
{Heat capacity measurements on FeAs-based compounds: a thermodynamic 
probe of electronic and magnetic states}

\author{P.\ J.\ Baker$^1$, S.\ R.\ Giblin$^2$, F.\ L.\ Pratt$^2$, 
R.\ H.\ Liu$^3$, G.\ Wu$^3$, X.\ H.\ Chen$^3$, M.\ J.\ Pitcher$^4$, 
D.\ R.\ Parker$^4$, S.\ J. Clarke$^4$, and S.\ J.\ Blundell$^1$}
\address{$^1$ Department of Physics, Oxford University, 
Parks Road, Oxford OX1 3PU, United Kingdom}
\address{$^2$ ISIS Muon Facility, Rutherford Appleton Laboratory, 
Harwell Science and Innovation Campus, Didcot, OX11 0QX, United Kingdom}
\address{$^3$ Hefei National Laboratory for Physical Sciences at Microscale 
and Department of Physics, University of Science and Technology of China, 
Hefei, Anhui 230026, China}
\address{$^4$ Department of Chemistry, University of Oxford, Inorganic 
Chemistry Laboratory, South Parks Road, Oxford, OX1 3QR, United Kingdom}

\ead{p.baker1@physics.ox.ac.uk}
\date{\today}

\begin{abstract}
We report heat capacity measurements of the pnictide materials 
SmFeAsO$_{1-x}$F$_x$, NdFeAsO, LaFeAsO$_{1-x}$F$_x$ and LiFeAs. For 
SmFeAsO$_{1-x}$F$_{x}$, with $x$ close to 0.1, we use $^3$He measurements to 
demonstrate a transfer of entropy from the peak at $T_{\rm N}$ to a previously 
unidentified $\sim 2$~K feature which grows with increasing doping. Our results 
on the Sm samples are compared with a similarly doped La sample to elucidate 
the crystal field levels of the Sm$^{3+}$ ion at $0$, $23$, and $56$~meV which 
lead to a Schottky-like anomaly, and also show that there is a significant 
increase in the Sommerfeld coefficient $\gamma$ when La is replaced by Sm 
or Nd. The lattice contribution to the heat capacity of the superconducting 
oxypnictides is found to vary negligibly with chemical substitution. 
% Not sure if this really should be included.
We also present a heat capacity measurement of LiFeAs showing the feature at 
$T_c$, which is significantly rounded and much smaller than the BCS value.
% Check this number. I think 25% would be more reasonable an estimate.
\end{abstract}

\submitto{NJP}
\maketitle

\section{\label{sec:intro} Introduction}

% Discovery of superconductivity in pnictide materials.
% Which groups of materials have been found to superconduct.
The discovery of superconductivity in pnictide compounds containing iron came as an 
enormous surprise to the condensed matter physics and materials chemistry community. 
This started with derivatives of LaFePO ($T_c=4$~K)~\cite{kamihara06jacs}, and then, 
attracting far greater attention, LaFeAsO$_{1-x}$F$_x$ ($T_c = 26$~K)~\cite{kamihara08jacs}. 
The {\it Ln}FeAsO$_{1-x}$F$_x$ (1111) materials have provided a remarkable range of 
superconducting materials with $T_c$ rising quickly to $43$~K in 
SmFeAsO$_{1-x}$F$_x$~\cite{chen08nature}, and subsequently $55$~K in 
that compound and oxygen deficient SmFeAsO$_{1-\delta}$~\cite{liu08prl,ren08epl}. 
Double layered FeAs-based (122) materials with the general formula 
$A$Fe$_{2}$As$_2$ ($T_c = 38$~K)~\cite{rotter08prl,sasmal08prl} and the single 
layered (111) and stoichiometric LiFeAs ($T_c \sim 16$K)~\cite{pitcher08,tapp08prb} 
and NaFeAs ($T_c = 9$~K)~\cite{parker08} also superconduct. 
%Selenium deficient \chem{FeSe} also superconducts

% What properties are of interest, and how heat capacity can tell us about them
% Superconducting properties, magnetism (both Fe and RE), nematic phase?
These new materials can be considered analogous to the cuprate high-$T_c$ 
superconductors but based on anti-PbO-type FeAs layers rather than planar CuO$_2$ 
layers and with more than one band relevant to the superconductivity. The 
superconducting properties and novel magnetism in these compounds have been a great 
spur to recent research. The tantalising prospect is that the analogy will offer 
insights that will identify common aspects of the superconducting mechanism in both 
types of material. Specific heat measurements are a powerful probe of magnetism and 
superconductivity since they probe the entropy involved with the electronic phase 
transitions, and provide a straightforward comparison with theoretical work. In 
the pnictide superconductors both Fe magnetism and superconductivity, and 
the magnetic behaviour of the rare-earth moments are amenable to investigation 
using this technique.

% Outline of paper
In this paper we present heat capacity results on three SmFeAsO$_{1-x}$F$_x$ samples 
in the doping range where the coexistence of magnetic order and superconductivity has been 
suggested, undoped SmFeAsO and NdFeAsO where magnetism is present without 
superconductivity, and superconducting samples of LaFeAsO$_{1-x}$F$_x$ and LiFeAs 
where there is no magnetic order. Following a review of previous work (section~2) and
a description of the technique (section~3), we present our results.  These are divided 
into the $^4$He heat capacity results on the oxypnictide compounds in section~\ref{sec:He4} 
and the $^3$He heat capacity results on the same materials in section~\ref{sec:He3}. The 
results on LiFeAs are presented in section~\ref{sec:LFA}. We describe the experimental 
methodology in section~\ref{sec:exp} and discuss the results in section~\ref{sec:discussion}.

\section{\label{sec:literature} Previous heat capacity studies of 
pnictide compounds}

A significant number of specific heat measurements have already been carried out on pnictide 
compounds although they have covered only a small selection of the compounds synthesized so 
far. SmFeAsO$_{1-x}$O$_{x}$ has already been studied at the dopings $x=0$, $0.05$, $0.07$, 
$0.15$, and $0.20$~\cite{ding08,tropeano08prb}, away from the intermediate doping range we 
consider here, revealing a peak in the specific heat in the undoped sample at 
$T_{\rm SDW} \sim 130$~K and $\lambda$ anomalies near $T_{\rm N} \sim 4$~K. 
In LaFeAs(O,F)~\cite{mcguire08prb,dong08epl,mu08cpl} the structural ($155$~K) and 
magnetic transitions ($143$~K) are clearly resolved and have a small combined entropy of 
$0.032$~R~\cite{dong08epl}. The Sommerfeld coefficient was found to be relatively small, 
$\gamma = 3.7$~mJmol$^{-1}$K$^{-2}$~\cite{dong08epl}. The low-temperature values of $\gamma$ 
were found to vary in applied magnetic field as $\gamma(H) \propto \sqrt{H}$, suggestive 
of line nodes in the superconducting gap function~\cite{mu08cpl}. In this issue 
McGuire {\it et al.}~\cite{mcguire08arxiv} examine the structural and magnetic transitions 
around $150$~K in CeFeAsO, PrFeAsO, and NdFeAsO, finding clear separation between them and 
distinct heat capacity anomalies at each transition. LaFePO has also 
been studied in detail~\cite{mcqueen08prb,kohama08arxiv,analytis08arxiv}, 
$\gamma \sim 10$~mJmol$^{-1}$K$^{-2}$ and relatively small specific heat jumps have been measured 
at $T_c$, $\Delta C_p / \gamma T_c \sim 0.6$.

A smaller number of 122 materials have also been measured: BaNi$_{2}$As$_{2}$ has a 
first-order phase transition at $T_o =130$~K and a Sommerfeld coefficient 
$\gamma = 10.8 \pm 0.1$~mJmol$^{-1}$K$^{-2}$. A superconducting sample with $T_c = 0.68$~K shows 
a very well defined specific heat jump at $T_c$, with $\Delta C /\gamma T_c = 1.31$. This 
is good evidence for bulk superconductivity, but further analysis was not 
pursued~\cite{ronning08jpcm}. 
A more recent study of a Ba$_{0.6}$K$_{0.4}$Fe$_{2}$As$_{2}$ single 
crystal~\cite{welp08arxiv} found a $\sim 1$~K wide specific heat peak 
at $T_c = 34.6$~K and probed the field dependence of the feature. 
Evidence for strong coupling was found and the $\kappa$ values were 
over $100$.  

There have been relatively few predictions made concerning the heat capacity of the pnictide 
superconductors, but one is particularly relevant to our work here~\cite{xu08prb}. It has been 
suggested that SmFeAsO$_{1-x}$F$_x$ has a nematic electronic phase overlapping the 
superconducting phase, probably in the doping range considered here. The low-temperature 
specific heat would change from $C \propto T^{\beta},~\beta < 2$ in the superconducting phase 
to $C \propto T^{2/3}$ in the nematic phase, with a crossover to the conventional $d=3$ scalings 
$C \propto T$ close enough to the nematic critical point. 

\section{\label{sec:exp} Experimental details}
% How the samples were made.
The SmFeAsO$_{1-x}$F$_{x}$ samples were synthesized by conventional solid state reaction 
methods, as described in Refs~\cite{chen08nature,liu08prl}. Standard powder x-ray diffraction 
patterns were measured where all peaks could be indexed to the tetragonal ZrCuSiAs-type 
structure. DC resistivity and magnetisation measurements were made to determine the midpoint 
(10~\% to 90~\% width) of the resistive and diamagnetic transitions with $T_c=10(7)$, $17(8)$, 
and $25(8)$~K for $x=0.1$, $0.12$, and $0.13$ respectively. These samples have previously been 
investigated using $\mu$SR~\cite{drew08arxiv}, which suggested the coexistence of superconductivity 
and magnetism. LaFeAsO$_{0.9}$F$_{0.1}$ ($T_c = 22$~K) was synthesised by a method adapted 
from Ref.~\cite{kamihara08jacs}; a mixture of La, La$_{2}$O$_{3}$, LaF$_3$, F 
and As was heated at $600^{\circ}$~C for 12~hrs in an evacuated silica ampoule then thoroughly 
re-ground, pelletised and heated to $1150^{\circ}$~C for a further 48hrs in evacuated silica. 
The resulting material was examined by laboratory powder x-ray diffraction and found 
to contain LaFeAsO$_{0.9}$F$_{0.1}$ ($T_c = 22$~K) as the majority phase with small amounts of FeAs 
($<5$mol~\%) and LaOF ($<2.5$mol~\%). 
The LiFeAs sample used in our experiments was synthesized by the method described in 
Ref.~\cite{pitcher08} and corresponds to Sample 2 described in that study, with $T_c = 12$~K. 
While the diamagnetic shift in this sample is far smaller than ideal, or that of the $T_c = 16$~K 
sample~\cite{pitcher08}, $\mu$SR measurements on this sample demonstrated that the superconducting 
sample volume was $> 80$~\%~\cite{pratt08arxiv}, suggesting that particle size effects may be 
affecting the bulk DC susceptibility measurement.

% How the experiments were done.
The heat capacity measurements were carried out using a Quantum Design Physical Properties 
Measurement System (PPMS) over the temperature range $0.4$ to $300$~K, with measurements below 
$2$~K carried out using the standard $^3$He insert. Samples were in the form of pressed powder pellets 
and were affixed to the measurement stage using Apiezon N-grease. The measurement technique 
employed by the PPMS is the two-tau relaxation method~\cite{hwang97,ppms00} where the temperature 
of the sample is measured while a heat pulse is applied and for an equal time afterwards while the 
sample cools. The time dependence of the temperature variation in each part of the measurement is 
described by the sum of two exponentials describing the heat capacity of the sample and the measurement 
stage, and the coupling between them. The effectiveness of this technique is discussed critically 
by Lashley~{\it et al.}~\cite{lashley03}, and we can expect the absolute heat capacity values to be 
correct to within $\pm 2$~\% for $5 \leq T \leq 300$~K and $\pm 5$~\% for $T \leq 5$~K. These errors 
are comparable with the error in the sample mass.

\section{\label{sec:He4} $^4$He results on oxypnictides}

\begin{figure}[htb]
\includegraphics[width=0.9\columnwidth]{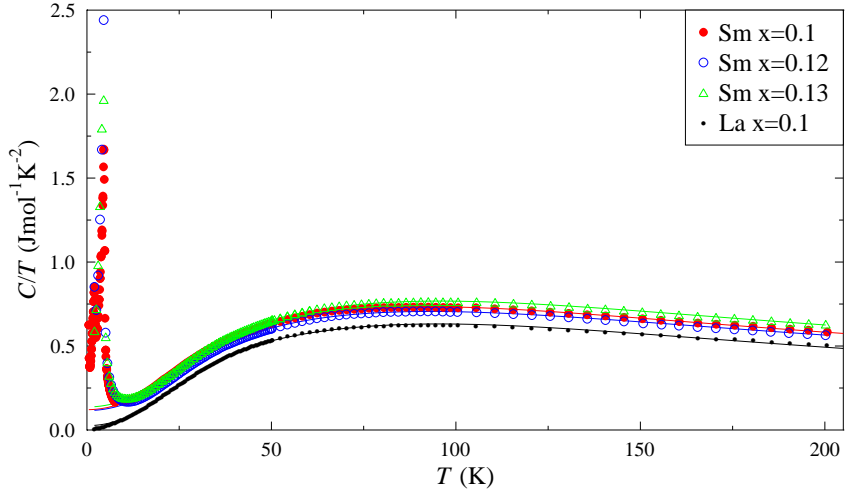}
\caption{\label{he4hc} 
% Edit this caption appropriately
Raw heat capacity data on \chem{SmFeAsO_{0.9}F_{0.1}}, 
\chem{SmFeAsO_{0.88}F_{0.12}}, \chem{SmFeAsO_{0.87}F_{0.13}}, and 
\chem{LaFeAsO_{0.9}F_{0.1}}. The fits are described in the text and 
use the parameters shown in Table~\ref{latticetable}. 
}
\end{figure}

The data for the oxypnictide materials SmFeAsO$_{0.9}$F$_{0.1}$, 
SmFeAsO$_{0.88}$F$_{0.12}$, SmFeAsO$_{0.87}$F$_{0.13}$, and 
LaFeAsO$_{0.9}$F$_{0.1}$ from $2$ to $200$~K are shown in Figure~\ref{he4hc}. 
The lattice and electronic parts of the heat capacity data are fitted to the equation:
\begin{equation}
C(T) = \gamma T + C_D (T) + C_E (T),
\label{eq:hcfit}
\end{equation}
where $\gamma$ is the Sommerfeld coefficient, and $C_D$ and $C_E$ are Debye and Einstein functions 
respectively. We found that the data were well described by including only one Einstein mode, giving a more simple 
fitting function than was used by Tropeano {\it et al.}~\cite{tropeano08prb}. The total amplitude of the lattice 
modes is almost exactly value expected from the Dulong-Petit law. Fits were carried out over the 
temperature range $10\leq T/{\rm K} \leq 200$~K as Apiezon N grease produces small heat capacity anomalies 
above $200$~K. The parameters derived from fitting equation~\ref{eq:hcfit} to the data shown in Figure~\ref{he4hc} 
are shown in Table~\ref{latticetable}. As is apparent from Figure~\ref{he4hc}, the lattice contributions are almost 
identical for the four samples, and the main differences are the change in the electronic term $\gamma$ going 
from the La compound to the Sm compounds, and the $\lambda$ anomalies at $T^{\rm Sm}_{\rm N}$ (hereafter $T_{\rm N}$) 
in each of the Sm samples. The difference between the $\gamma$ values of the Sm and La compounds, 
$\sim 120$ vs.~$26$~mJmol$^{-1}$K$^{-2}$, is quite remarkable. Our values of $\gamma$ are comparable with those reported 
by Ding {\it et al.}~\cite{ding08} for samples with slightly different dopings. Similarly high values of $\gamma$ 
have been observed in Nd (as we show in section~\ref{sec:He3} below) and Ce oxypnictides. 
% Must check exact details of this. Certainly on a poster at TEM08. Note our value of gamma for Nd.

% Need I add anything here.

\begin{table}[b]
\caption{\label{latticetable} Lattice heat capacity fitting parameters derived from fitting 
Equation~\ref{eq:hcfit} to the raw heat capacity data for the SmFeAsO$_{1-x}$F$_{x}$ and 
\chem{LaFeAsO_{0.9}F_{0.1}} samples described in Section~\ref{sec:He4}. The quoted errors 
are $1 \sigma$ from the fitting.  
}
%\begin{center}
\begin{tabular}{ l| c| c| c | c}
\hline
Doping ($x$) & Sm 0.1 & Sm 0.12 & Sm 0.13 & La 0.1  \\
\hline
$\theta_{\rm D}$ (K) &  212(1) &  227(2)  & 227(2)  & 212(1) \\
$\theta_{\rm E}$ (K) &  380(3) &  404(13)  & 407(8) & 394(4)  \\
$A_{\rm D}$ (Jmol$^{-1}$K$^{-2}$) & 51(1) & 55(1) & 58(1) & 51(1) \\
$A_{\rm E}$ (Jmol$^{-1}$K$^{-2}$) & 59(1) & 53(2) & 60.9(1) & 62(1) \\
$\gamma$ (mJmol$^{-1}$K$^{-2}$) & 121(1) & 116(1) & 137(1) & 26(7) \\
\hline
\end{tabular}
%\end{center}
\end{table}

\begin{figure}[htbp]
\includegraphics[width=0.9\columnwidth]{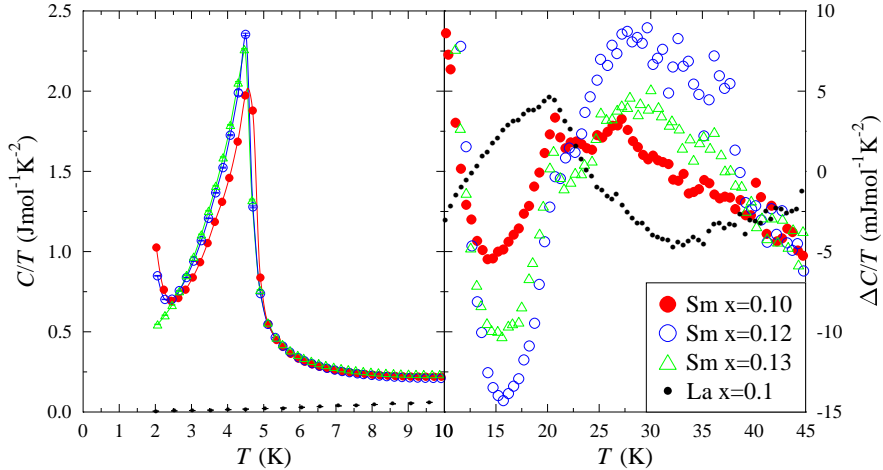}
\caption{\label{LnTNTc}
(Left) $C/T$ around $T_{\rm N}$ showing the shape of the peaks, and the upturn at low 
temperatures which we investigate further in section~\ref{sec:He3}. (Right) Difference between 
the data and the lattice + linear fit in the region of $T_c$ for each of the samples. In 
\chem{LaFeAsO_{0.9}F_{0.1}} the peak can be identified with the superconducting transition, whereas in the 
three SmFeAsO$_{1-x}$F$_{x}$ samples \chem{Fe} magnetic ordering also contributes.
}
\end{figure}

In Figure~\ref{LnTNTc} the raw data near $T_{\rm N}$ and the residues 
between the data in Figure~\ref{he4hc}  
and the fits given in Table~\ref{latticetable} near $T_c$ are shown. The 
peak in $C/T$ near $T_{\rm N}$ is clear for each of the samples and we 
can see the slow drop in $T_{\rm N}$ with increasing doping. The height 
of the peak also seems to drop with increasing doping. Carrying out 
entropy subtractions suggested that the magnetic entropy was also dropping 
as the doping was increased and in the $x=0.1$ and $0.12$ samples an 
upturn in $C/T$ was observed at the lowest temperatures, though as we 
will see in section~\ref{sec:He3} this is a hint of what is actually going 
on at low-temperatures. 

Near the superconducting $T_c$ and $T^{\rm Fe}_{\rm N}$ in the right-hand 
panel of Figure~\ref{LnTNTc} the difference between the data and the fits 
is far smaller. It seems that the peak around $T_c$ in the \chem{La} 
compound is well defined and of a similar size to those found 
previously, though somewhat broader. In the \chem{Sm} compounds the temperatures 
of $T_c$ and $T^{\rm Fe}_{\rm N}$ have been determined by other techniques~\cite{drew08arxiv}. 
The feature seems to grow with increasing doping and in the $x=0.1$ sample must correspond 
to \chem{Fe} ordering. In the $x = 0.12$ sample the anomaly starts at $T_c$ and 
continues up to $T^{\rm Fe}_{\rm N}$. The $x=0.13$ sample has almost coincident 
$T_c$ and $T^{\rm Fe}_{\rm N}$ and we observe a broad and less structured feature 
at the appropriate temperature. 

Seeing that the entropy values extracted from the low-temperature N{\'{e}}el ordering of 
the \chem{Sm} magnetic moments approximate much more closely to $S=1/2$ than the $S=5/2$ 
expected for free ions~\cite{cimberle08arxiv} we sought to investigate the possibility of 
a Schottky-like anomaly due to the crystal field levels being gradually populated with 
increasing temperature. This can be done by comparing the heat capacity of \chem{Sm} and 
\chem{La}-containing oxypnictides, but it would be challenging to attempt this on undoped 
samples since significant anomalies appear at the structural and magnetic phase transition(s) 
near $150$~K. Instead, we used our measurements of the $10$~\% doped samples where the 
superconducting and magnetic anomalies above $10$~K are very small. From the \chem{La} 
compound we get an excellent estimate of the lattice contribution because there is no 
rare-earth moment, and we can easily adjust for the difference in the Sommerfeld coefficient 
in our fitting of the difference. We took a direct molar subtraction between data sets 
measured at common temperature points and fitted the difference, 
$\Delta C=C(\chem{Sm})-C(\chem{La})$, to the function:
\begin{eqnarray}
\Delta C &=& \Delta \gamma + R \left(\frac{Z_2}{Z_0}-\frac{Z_1}{Z_0} \right)\left(\frac{Z_1}{Z_0}\right),
\label{eq:doubleschottky}\\
Z_0 &=& g_1 \exp\left(-\frac{\Delta_1}{T}\right) + g_2 \exp\left(-\frac{\Delta_2}{T}\right), \label{Z0} \\
Z_1 &=& \frac{g_1 \Delta_1}{T} \exp\left(-\frac{\Delta_1}{T}\right) + \frac{g_2 \Delta_2}{T} 
\exp\left(-\frac{\Delta_2}{T}\right), \label{Z1} \\
Z_2 &=& \frac{g_1 \Delta^{2}_1}{T^2} \exp\left(-\frac{\Delta_1}{T}\right) + 
\frac{g_2 \Delta^{2}_2}{T^2} \exp\left(-\frac{\Delta_2}{T}\right), \label{Z2}  
\end{eqnarray}  
where $\Delta \gamma$ is the difference between the electronic specific heats, $g_i$ are the 
degeneracies of each of the levels,  and $\Delta_1$ and $\Delta_2$ are the energy gaps (in K) 
of the $S=1/2 \rightarrow S=3/2$ and $S=1/2 \rightarrow S=5/2$ transitions respectively.
The parameters derived from this analysis are $\Delta_1 = 266(3)$~K~$=22.92(3)$~meV and 
$\Delta_2 = 650(50)$~K~$=56(4)$~meV. The former value is in agreement with the estimates 
coming from the thermal activation of the $\mu$SR relaxation rate~\cite{drew08prl,khasanov08prb} 
and the latter value has previously been derived from magnetic susceptibility 
measurements ($\Delta = 520$~K for $x=0$ and $\Delta = 620$~K for $x=0.15$)~\cite{cimberle08arxiv}.
\chem{Sm} containing materials are not amenable to investigation using inelastic neutron scattering, 
which is generally the most powerful probe of the crystal field levels in these materials. However, 
CeFeAsO$_{1-x}$F$_{x}$ has been investigated, and crystal field doublets at $0$, $18.7$, and 
$58.4$~meV have been found for an $x=0.16$ superconducting sample~\cite{chi08arxiv}. In the 
undoped sample three doublets in the paramagnetic phase split into six singlets in the antiferromagnetic 
phase. It is interesting to note the similarity between the crystal field levels in these two materials and 
the observation that the $18.7$~meV mode has a strongly enhanced local magnetic susceptibility below 
$T_c$, which is reminiscent of the increased $\mu$SR relaxation rates observed for the \chem{Sm} 
compounds~\cite{drew08prl,khasanov08prb}. 
The other parameter extracted from this analysis is the difference in the Sommerfeld coefficients, 
$\Delta\gamma = 53(2)$~mJmol$^{-1}$K$^{-2}$, which is far lower than the values extracted from fitting the raw data. 
This would suggest that the true value of $\gamma \sim 80$~mJmol$^{-1}$K$^{-2}$ in the \chem{Sm} oxypnictides - 
still significantly larger than for \chem{LaFeAs(O,F)}. (Ding {\it et al.}~\cite{ding08} found a similar 
value from some of their fitting.) It seems likely that failing to include the Schottky-like anomaly due 
to the crystal field levels in the fitting of the raw data leads to an erroneously high value of $\gamma$, 
as we found above and has been noted by other reports~\cite{ding08}. 
Observing the doping dependence of the crystal field levels in these materials is likely to prove 
a fruitful area of future study, particularly since the rare-earth moment fluctuations are evidently 
well coupled to the \chem{Fe} fluctuations in the \chem{FeAs} planes~\cite{drew08prl,chi08arxiv}.

\section{\label{sec:He3} $^3$He results on oxypnictides}

\begin{figure}[htb]
\includegraphics[width=0.9\columnwidth]{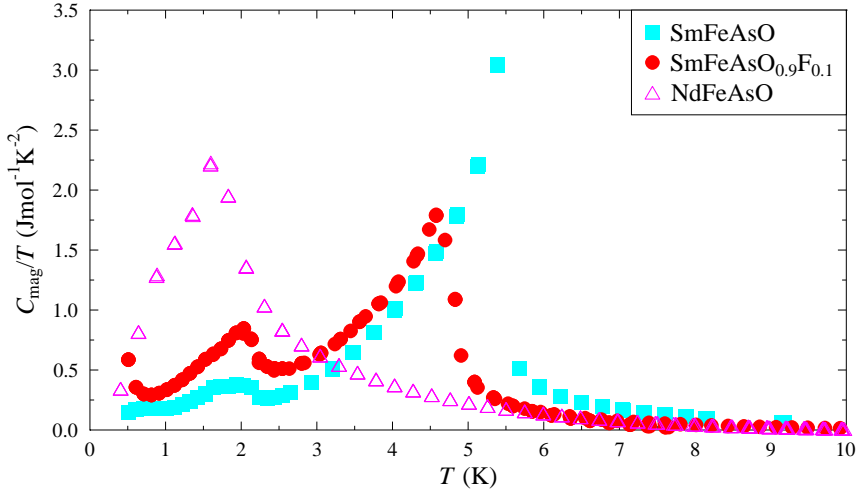}
\caption{\label{He3} 
$^3$He heat capacity data on \chem{SmFeAsO}, \chem{NdFeAsO}, and \chem{SmFeAsO_{0.9}F_{0.1}}. 
The subtraction of lattice terms to calculate $C_{\rm mag}/T$ is discussed in the text. 
% Expand this caption
}
\end{figure}

Our $^3$He heat capacity results for \chem{SmFeAsO}, \chem{SmFeAsO_{0.9}F_{0.1}}, and \chem{NdFeAsO} are plotted 
in Figure~\ref{He3} after the subtraction of non-magnetic backgrounds. We made the assumption that the lattice 
contribution to the \chem{Sm} samples would be identical and used the \chem{La} lattice background for the \chem{Nd} 
and corrected for the difference in the Sommerfeld coefficients, estimating a value for the \chem{Nd} compound of 
$\sim 60$~mJmol$^{-1}$K$^{-2}$. The \chem{Nd} and \chem{Sm} magnetic ordering leads to well defined $\lambda$ anomalies 
at the N\'{e}el temperature of the rare-earth moments. We find $T^{\rm Nd}_{\rm N} = 1.9$~K in reasonable agreement 
with the value determined by neutron diffraction ($T_{\rm N} = 1.96(3)$~K~\cite{qiu08arxiv}). Other neutron 
diffraction measurements have shown that the \chem{Fe} moments order at around $141$~K with an ordered moment of 
$0.25(7)~\mu_{\rm B}$~\cite{chen08prb}. We discuss the temperature variation of $T^{\rm Sm}_{\rm N}$ with doping 
in section~\ref{sec:discussion}.

In the two \chem{Sm} samples we see a further anomaly below the previously determined $T_{\rm N}$ values at around 
$2$~K. This feature grows significantly with increasing $x$ as the feature at $T_{\rm N}$ decreases in amplitude. We 
are unaware of any previous measurements on SmFeAsO$_{1-x}$F$_{x}$ below $2$~K that have shown this feature, though 
our $^4$He measurements show an upturn at low-temperature (Figure~\ref{LnTNTc}) that implies something must be 
occurring, and the decrease in the height at $T_{\rm N}$ has been reported elsewhere, if not 
discussed~\cite{ding08,tropeano08prb}. We note that if the superconducting state is becoming more robust with 
increasing doping there will be a drop in the electronic contribution to the specific heat below $T_c (> T_{\rm N})$. 
This would lead to a drop in the height of the magnetic features before the background is subtracted and a reduced 
estimate of the magnetic entropy after background subtraction. 
% Need to put in an explanation or at least hypothesis for this anomaly.

\section{\label{sec:LFA} \chem{LiFeAs}}

\begin{figure}[htb]
\includegraphics[width=0.9\columnwidth]{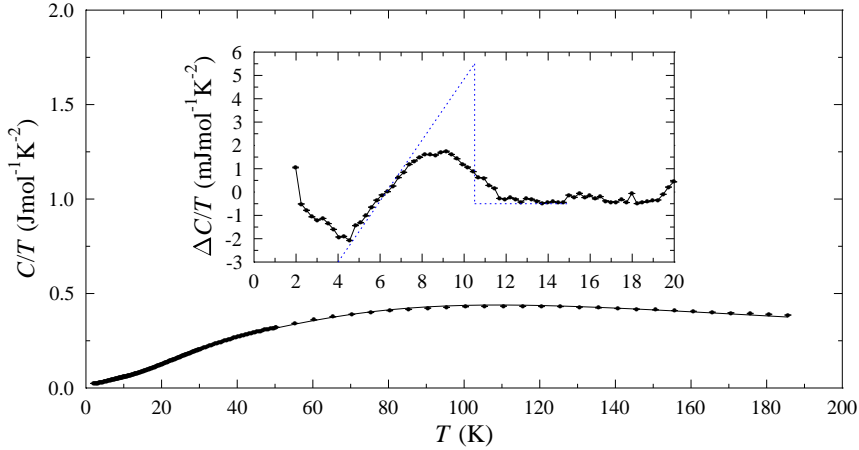}
\caption{\label{LFA} 
Heat capacity of \chem{LiFeAs}. The inset shows the difference between 
$0$ and $12$~T points, with a triangular approximation to the peak discussed
in the text. 
}
\end{figure}

The \chem{LiFeAs} sample ($T_c = 12$~K) gives us an opportunity to observe the superconducting behaviour without 
the effects of magnetic ordering of either rare-earth or iron moments. Since it superconducts without any 
non-stoichiometry we compare our results with those on \chem{LaFePO} reported 
elsewhere~\cite{mcqueen08prb,kohama08arxiv,analytis08arxiv}. 
Data on \chem{LiFeAs} in zero-field and $12$~T are shown in Figure~\ref{LFA} and we plot the difference 
between them in the inset. A clear peak is shown in the vicinity of $T_c$ although it is evidently broadened, as 
we would expect on the basis of susceptibility measurements~\cite{pitcher08}.
(We also see a difference between the heat capacity traces above $20$~K, which we cannot explain.) 
% Can we explain this, and do we have to mention it?
In the inset we plot a linear trend following the low-temperature behaviour of the peak in the residue and 
extending up to the midpoint of the transition to estimate the size of the specific heat anomaly in this compound. 
This has two inherent difficulties: the transition is relatively broad and may affect the curvature below $T_c$ 
to a greater extent than we have assumed, and more importantly, the upper critical field, $B_{c2}$, is far greater 
than our measurement field. On the basis of susceptibility measurements $B_{c2}>80$~T~\cite{tapp08prb}, 
a result which is consistent with the value extracted from $\mu$SR measurements~\cite{pratt08arxiv}. $T_c$ is therefore  
likely to have only been shifted by $<1$~K by our measurement field and so the residue does not have a full 
subtraction of the superconducting part of the heat capacity. 
From the line plotted in the inset we can estimate the normalized specific heat jump at $T_c$ to be 
$\Delta C /\gamma T_c = 0.26$, with $\gamma \sim 23$~mJmol$^{-1}$K$^{-2}$. This is about $1/3$ the value found for 
\chem{LaFePO}, which has a much lower $B_{c2} \sim 3.5$~T~\cite{analytis08arxiv}. 
The small specific heat jump might suggest a small superconducting sample volume, which magnetic susceptibility 
measurements also suggested~\cite{pitcher08}, but $\mu$SR measurements on a sample from the same batch gave 
a superconducting sample volume $>80$~\%~\cite{pratt08arxiv}. On this basis the small shift in $T_c$ is more 
likely to have led to the small specific heat jump. Both higher applied magnetic fields and samples with a more 
sharply defined $T_c$ will be necessary to gain a more complete understanding of the thermodynamic properties 
of \chem{LiFeAs}.

\section{\label{sec:discussion} Discussion}

Our results show a new feature in the heat capacity of SmFeAsO$_{1-x}$F$_x$ at around $2$~K 
which appears to grow and shift to lower temperature with increasing doping. We are not aware of 
this feature being reported elsewhere, nor any predictions of such a feature. Since the entropy 
released at the $T_{\rm N}$ peak seems to be dropping as this feature grows it seems likely that 
this $2$~K feature is related to the \chem{Sm} moments. The presence of the feature at low-temperature 
in the undoped sample suggests that it is not related to the nematic phase suggested by Xu 
{\it et al.}~\cite{xu08prb} if their phase diagram is appropriate. Alternatives may be successive 
ordering of components of the \chem{Sm} magnetic moments or reorientations of the moments between states 
with a different entropy, which are altered by the doping-dependent orthorhombic to tetragonal structural 
distortion~\cite{margadonna08}. 
In \chem{NdFeAsO} we have verified the $T_{\rm N}$ value previously determined using neutron 
scattering~\cite{qiu08arxiv}. In Table~\ref{magnetictable} we have collated the values of $T_{\rm N}$ 
determined in this study and the magnetic entropy values extracted. Our undoped \chem{Sm}-oxypnictide 
has a higher $T_{\rm N}$ than has been suggested by previous studies~\cite{ding08,tropeano08prb}. It is 
not clear at this stage why this value is increased above previous measurements, although our value does sit 
closer to the trend of decreasing $T_{\rm N}$ with increasing doping suggested by other measurements. 
Where we have $^3$He data the magnetic entropy values are $R\ln{2}$ within the experimental error (we 
take this to be that of the low temperature accuracy of the cryostat) but for the \chem{SmFeAsO_{0.88}F_{0.12}} 
and \chem{SmFeAsO_{0.87}F_{0.13}} the data above $2$~K suggest a significant amount of entropy is missing. 
Given the behaviour seen in the \chem{SmFeAsO_{0.9}F_{0.1}} sample, it seems very likely that a 
similar feature is present below $2$~K and accounts for this missing entropy. 
The entropy values we obtain for the rare-earth moment ordering suggest that both \chem{Sm} and 
\chem{Nd} are in a $S=1/2$ state at low-temperature. In the \chem{Sm} $x=0.1$ sample we found 
that the higher temperature heat capacity includes a Schottky-like anomaly due to the higher 
crystal field levels at $23$ and $56$~meV above the ground state. These levels are similar to 
those observed for superconducting \chem{CeFeAsO_{0.84}F_{0.16}} using inelastic neutron scattering 
($0$, $18.7$, and $58.4$~meV)~\cite{chi08arxiv}. It will be interesting to discover the variation of 
these crystal field levels in other materials as this field of research progresses, particularly if the 
rare-earth moments have any influence on the superconductivity.

The heat capacity features associated with the superconducting transitions in the SmFeAsO$_{1-x}$F$_x$ 
are less well defined than those reported for superconducting samples without any magnetic transitions. 
The fit residues suggest that both magnetic and superconducting transitions contribute features, but while they 
are present at the temperatures suggested by other techniques~\cite{drew08arxiv}, the entropy change is small 
and diffuse. The features in \chem{LaFeAsO_{0.9}F_{0.1}} and \chem{LiFeAs} are much better defined and are 
comparable with other reports on similar samples.

\begin{table}[b]
\caption{\label{magnetictable} 
N\'{e}el temperatures and magnetic entropy values for the samples investigated in this work. The 
methodology is discussed in Sections~\ref{sec:He4}~and~\ref{sec:He3}. Entropies extracted for the 
\chem{SmFeAsO_{0.88}F_{0.12}} and \chem{SmFeAsO_{0.87}F_{0.13}} are estimates due to the lack of 
$^3$He data, calculated assuming a smooth trend down to $T=0$. The error bars are based on the quoted 
accuracy of the calorimeter at low temperatures~\cite{ppms00,lashley03}.
}
%\begin{center}
\begin{tabular}{ l| c| c}
\hline
Doping ($x$) & $T_{\rm N}$ (K) & $S_{\rm mag}/R\ln{2}$  \\
\hline
\chem{NdFeAsO} &  1.90(5) &  0.96(10) \\
\chem{SmFeAsO} &  5.4(1) &  0.91(10) \\
\chem{SmFeAsO_{0.9}F_{0.1}} &  4.65(5) &  0.95(10) \\
\chem{SmFeAsO_{0.88}F_{0.12}} &  4.50(5) &  $\sim 0.90$ \\
\chem{SmFeAsO_{0.87}F_{0.13}} &  4.45(5) &  $\sim 0.67$ \\
\hline
\end{tabular}
%\end{center}
\end{table}

\section{\label{sec:conc} Conclusions}
% Concluding paragraph
% What we have found out
% What we did that was best
We have studied a series of Sm, Nd, and La-containing oxypnictides and investigated the effect that 
the rare-earth magnetic moments have on the magnetism and superconductivity. The $\lambda$ anomalies 
at each N\'{e}el temperature give clear evidence for ordering of the rare-earth moments, confirming 
previous studies. We have extended those studies by investigating the form of the heat capacity in 
the Sm samples below 2~K, which provides evidence for a new feature occurring slightly below 2~K 
that grows with increasing doping. This provides a natural explanation for the previously observed 
drop in the size of the peak at $T_{\rm N} \sim 4$~K with increasing doping. We also probed the 
crystal field levels of the Sm$^{3+}$ ions in SmFeAsO$_{0.9}$F$_{0.1}$ and found they were consistent 
with both the $R\ln{2}$ entropy associated with magnetic ordering at low-temperature and also the 
free spin $S=5/2$ value observed at high temperature. The energies of the levels appear to be 
similar to those found in superconducting \chem{CeFeAsO_{0.84}F_{0.16}} using inelastic neutron scattering. 
In LaFeAsO$_{0.9}$F$_{0.1}$ and \chem{LiFeAs} we see distinct anomalies due to the superconducting 
transitions, whereas in the SmFeAsO$_{1-x}$F$_x$ samples magnetic contributions complicate the 
form of the data and the superconducting contribution is less distinct. There remain many avenues to 
be explored using heat capacity in this fast moving field, including the possibility 
of nematic electronic order when superconductivity has been suppressed in SmFeAsO$_{1-x}$F$_x$ and 
the trends of crystal field levels as the rare-earth ions are substituted and the oxygen doping is changed; 
but it is already clear that this fundemental thermodynamic technique, coupled with careful analysis, 
can provide much useful information on the properties of FeAs-based compounds.

\ack{
We are grateful to Christian Bernhard, Andrew Boothroyd, Alan Drew, and Paul Goddard for helpful 
discussions,  and to Prabhakaran Dharmalingam for experimental assistance. 
This work was supported by the EPSRC and STFC (UK).
}
\\
% Check order of references

\end{document}